\documentclass[superscriptaddress,twocolumn,showpacs,prl,floatfix]{revtex4}

\usepackage{epsfig}
\usepackage{times}
\begin{document}

\noindent {\bf Surer {\em et al}.~Reply:} In the previous Comment
\cite{moebius:09}, M\"obius and Richter (MR) criticize the
secondary result of our Letter \cite{surer:09-ea} because of our
finite-size scaling (FSS) ansatz for the density of states of
the three-dimensional Coulomb glass (CG). Thus, our main result,
namely that the CG does not exhibit a phase transition down to the
lowest numerically accessible temperatures, remains unchallenged.
This finding suggests an absence of a finite-temperature CG phase --
an issue that has remained unsettled in the past 35 years and that
has been independently verified recently \cite{goethe:09}.  However,
we state that the third result of our Letter, namely, the density of
state (DOS) data for the random displacement (RD) version of the model,
is incorrect.

In contrast to the lack of conclusive results on the existence of
a phase transition, there have been several studies attempting to
verify the estimate of Efros and Shklovskii (ES) \cite{efros:75}
that the exponent of the DOS is exactly $\delta = D - 1$, where $D$
is the space dimension. One central question of
the field is whether numerical studies verify or contradict the
ES prediction.  As discussed in Ref.~\cite{surer:09-ea} there are many
(often nonoverlapping) predictions on the exponent $\delta$. Despite
these differences, most publications agree that $\delta$ is different
from the ES prediction to some degree and that it seems to be disorder
dependent.

In an attempt to settle the controversy regarding the exponent
$\delta$, we evaluated the DOS of the CG by identifying states as close
as possible to the exact ground state with the extremal optimization
(EO) method \cite{boettcher:05}.  Our results using EO, however,
did not deliver results qualitatively different from the published
ones: $\delta$ is different from the ES prediction to some degree
and apparently disorder dependent.  MR's own work \cite{moebius:92-ea}
also found these attributes for $\delta$.

In Ref.~\cite{surer:09-ea} we write that ``the DOS can be fit very
well with a form $\sim |E|^\delta$'' with $\delta$ ``close to the ES
value'' of $2$ and not {\em equal to $2$}.  Our estimate differs up to
$8$\%--$10$\% from the ES prediction, whereas MR's \cite{moebius:92-ea}
estimates are $10$\%--$25$\% off.  That MR prefer to describe this
difference as a ``clear deviation,'' whereas we used the words
``close to,'' is an insubstantial matter of semantics.

MR criticize our FSS form for the DOS, $\rho(E,N) \sim b
+ a|E|^\delta$.  Here $E$ is the energy, $N = L^D$ is the number
of sites with $L$ the linear dimension of the sample, and $b(N) =
\rho(E=0,N)$.  To our knowledge, there have been no attempts to perform
a {\em clean} FSS of the numerical data beyond the unmotivated (albeit
conservative) $1/L$ cutoff for the pure power-law fitting function
in Ref.~\cite{moebius:92-ea}. The latter corresponds to a step function
in the FSS form, something that could be deemed as unphysical. This
is mirrored by the large cutoff-dependent variations of the exponents
when performing fits to the pure power laws \cite{moebius:92-ea}. To avoid
such artifacts, we chose to smoothly connect the low- and high-energy
asymptotic behavior with the aforementioned formula.  We do, however,
agree with MR that the finite-size effects of $\rho(E=0,N)$ are indeed
puzzling and require further study; i.e., simulations of much larger
systems would be needed to verify the $b(N)$ dependence.

Although MR's results are qualitatively similar to ours and bring no
new insights to the problem we outline some details of their numerical
calculations that prevent a fair comparison of both data sets. First,
MR neglect the long-ranged nature of the Coulomb interactions by
using the minimal distance method. Unphysical edge effects arise
without the use of resummation methods that can influence the
thermodynamic behavior. Second, MR propose that their data differ
from ours because the ``minimum search [we used] does not consider
all single-particle hops.'' This is not correct because EO allows for
the rearrangement of any number of particles with hops of arbitrary
length. And indeed, we verified that our results are stable against
single-particle rearrangements. We note that EO was not as efficient
as we expected from studies on the Sherrington-Kirkpatrick spin glass
\cite{boettcher:05}.

We have also analyzed the discrepancies for the RD model.  We reviewed
our work and found an error in our code. Therefore, the displayed DOSs
\cite{surer:09-ea} correspond to disorder values that are $\sim 10$
times smaller than indicated.  We have repeated the RD simulations
with the corrected code, as well as with a new independent code based
on single-particle hops. The DOSs obtained with the two methods are
consistent and reveal that the Wigner crystal is not as robust against
disorder as stated in Ref.~\cite{surer:09-ea}.

Summarizing, the Comment does not offer new physical insights.  We
correct our statements about the RD model in Ref.~\cite{surer:09-ea}:
The Wigner crystal is not as robust to disorder as stated.  Still,
our main result of a lack of a finite-temperature transition
in the CG remains unchallenged.

\smallskip

\noindent B.~Surer,$^1$ A.~Glatz,$^2$ H.~G.~Katzgraber,$^{3,1}$
G.~T.~Zimanyi,$^4$ B.~A.~Allgood,$^5$ and G.~Blatter$^1$

\smallskip

\noindent $^1$Theoretische Physik, ETH Zurich, CH-8093 Zurich,
Switzerland; $^2$Materials Science Division, Argonne National
Laboratory, Argonne, Illinois 60439, USA; $^3$Department of Physics and
Astronomy, Texas A\&M University, College Station, Texas 77843-4242,
USA; $^4$Department of Physics, University of California, Davis,
California 95616, USA; $^5$Numerate Inc., San Bruno, California
94066, USA

\smallskip


\vspace*{-0.5em}

\bibliography{refs}

\begin{thebibliography}{6}
\expandafter\ifx\csname natexlab\endcsname\relax\def\natexlab#1{#1}\fi
\expandafter\ifx\csname bibnamefont\endcsname\relax
  \def\bibnamefont#1{#1}\fi
\expandafter\ifx\csname bibfnamefont\endcsname\relax
  \def\bibfnamefont#1{#1}\fi
\expandafter\ifx\csname citenamefont\endcsname\relax
  \def\citenamefont#1{#1}\fi
\expandafter\ifx\csname url\endcsname\relax
  \def\url#1{\texttt{#1}}\fi
\expandafter\ifx\csname urlprefix\endcsname\relax\def\urlprefix{URL }\fi
\providecommand{\bibinfo}[2]{#2}
\providecommand{\eprint}[2][]{\url{#2}}

\bibitem[{\citenamefont{{M{\"o}bius} and {Richter}}(2009)}]{moebius:09}
\bibinfo{author}{\bibfnamefont{A.}~\bibnamefont{{M{\"o}bius}}}
  \bibnamefont{and} \bibinfo{author}{\bibfnamefont{M.}~\bibnamefont{{Richter}}}
  (\bibinfo{year}{2009}), \bibinfo{note}{preceding Comment (see also
  arXiv:0908.3092)}.

\bibitem[{\citenamefont{Surer~{\em et al.}}(2009)}]{surer:09-ea}
\bibinfo{author}{\bibfnamefont{B.}~\bibnamefont{Surer~{\em et al.}}},
  \bibinfo{journal}{Phys. Rev. Lett.} \textbf{\bibinfo{volume}{102}},
  \bibinfo{pages}{067205} (\bibinfo{year}{2009}).

\bibitem[{\citenamefont{Goethe and Palassini}(2009)}]{goethe:09}
\bibinfo{author}{\bibfnamefont{M.}~\bibnamefont{Goethe}} \bibnamefont{and}
  \bibinfo{author}{\bibfnamefont{M.}~\bibnamefont{Palassini}},
  \bibinfo{journal}{Phys. Rev. Lett.} \textbf{\bibinfo{volume}{103}},
  \bibinfo{pages}{045702} (\bibinfo{year}{2009}).

\bibitem[{\citenamefont{{Efros} and {Shklovskii}}(1975)}]{efros:75}
\bibinfo{author}{\bibfnamefont{A.~L.} \bibnamefont{{Efros}}} \bibnamefont{and}
  \bibinfo{author}{\bibfnamefont{B.~I.} \bibnamefont{{Shklovskii}}},
  \bibinfo{journal}{J. Phys. C} \textbf{\bibinfo{volume}{8}},
  \bibinfo{pages}{L49} (\bibinfo{year}{1975}).

\bibitem[{\citenamefont{Boettcher}(2005)}]{boettcher:05}
\bibinfo{author}{\bibfnamefont{S.}~\bibnamefont{Boettcher}},
  \bibinfo{journal}{E. Phys. J. B} \textbf{\bibinfo{volume}{46}},
  \bibinfo{pages}{501} (\bibinfo{year}{2005}).

\bibitem[{\citenamefont{{M{\"o}bius}~{\em et al.}}(1992)}]{moebius:92-ea}
\bibinfo{author}{\bibfnamefont{A.}~\bibnamefont{{M{\"o}bius}~{\em et al.}}},
  \bibinfo{journal}{Phys. Rev. B} \textbf{\bibinfo{volume}{45}},
  \bibinfo{pages}{11568} (\bibinfo{year}{1992}).

\end{thebibliography}

\end{document}